\documentclass[a4paper, 12pt]{article}
\usepackage[dvips]{graphicx}

\begin{document}

\title{Chiral symmetry breaking and stability of strangelets}
\author{S. Yasui, A. Hosaka and H. Toki \\
{\it Research Center for Nuclear Physics (RCNP), Osaka University} \\
{\it 10-1 Mihogaoka, Ibaraki, Osaka, 567-0047, Japan} }
\maketitle

\begin{abstract}
We discuss the stability of strangelets by considering dynamical chiral symmetry breaking and confinement.
We use a $U(3)_{L}\!\times\!U(3)_{R}$ symmetric Nambu--Jona-Lasinio model for chiral symmetry breaking supplemented by a boundary condition for confinement.
It is shown that strangelets with  baryon number $A\!<\!2\!\times\!10^{3}$ can stably exist.
For the observables, we obtain the masses and the charge-to-baryon number ratios of the strangelets.
These quantities are compared with the observed data of the exotic particles.
\end{abstract}

\section{Introduction}

The strange matter, containing the $u$, $d$ and $s$ quarks,  has been considered to be the ground state of QCD \cite{Bodmer, Witten, Fahri}, and expected to play an important role in the astrophysical phenomena such as the quark stars and the early universe.
It is also interesting that droplets of the strange matter (strangelets) could be a candidate of the dark matter.
There is also a possibility to observe strangelets in the relativistic heavy ion collisions.

Whereas the physics of the quark matter is interesting in many respects, it is difficult to describe its properties directly from QCD.
Until now, the stability of the strange matter has been discussed by using effective models of QCD.
In the early stage, 
the MIT bag model was often used with an assumption that the strange matter could be treated as a system of free fermi gas in a bag 
\cite{Bodmer, Witten, Fahri, Madsen}.
There, the strange matter became stable than the $ud$ quark matter due to the large number of the degrees of freedom by including the strangeness.
In these works, not only the bulk quark matter, but also the strangelet of finite size has also been studied.
It was then shown that the strangelets could be more stable than the normal nuclei. 

However, in the MIT bag model, an important feature of QCD is not incorporated, that is the dynamical chiral symmetry breaking.
In fact, when it is taken into account, it has been shown that the strange matter cannot be absolutely stable \cite{Buballa96, Buballa98, Buballa99}.
The pattern of chiral symmetry restoration in the finite density quark matter is different for the $ud$ quarks and the $s$ quarks.
The chiral symmetry for the $ud$ quarks is sufficiently restored at stable densities ($n_{B}\!\simeq\!2\!-\!3n_{B}^{0}$),  while that for the $s$ quarks is still largely broken.
Then, the transition from the $ud$ quark matter to the strange matter by weak process is disfavored because of the large dynamical quark mass in the strange quark sector.
This result is qualitatively very different from the results of the absolute stability obtained in the MIT bag model, in which only current quark masses are used.

Though the strange matter of infinite volume was discussed  by taking into account dynamical chiral symmetry breaking \cite{Buballa96, Buballa99, Buballa98}, the strangelet with finite volume is not discussed yet.
In this paper, we study the stability of the strangelet by considering dynamical chiral symmetry breaking. 
In section 2, 
we formulate the lagrangian to describe dynamical chiral symmetry breaking supplemented by the confinement which is treated approximately by a boundary condition.
In section 3, numerical results are shown and we discuss the stability of the strangelet and present several observables.
In section 4, we conclude our discussions.


\section{Formulation}

We consider the NJL model with a four-point quark interaction  \cite{Nambu}.
This interaction is responsible for the dynamical generation of the quark mass.
For a finite system of strangelet, we also have to consider that quarks are confined in a cavity with finite volume.
To incorporate the latter aspect, we introduce the boundary condition of the MIT bag model.
Therefore, our model lagrangian for the strangelet is given by \cite{Kim, Kiriyama_Hosaka}
\begin{eqnarray}
{\cal L} \!=\! \bar{\psi}(i\partial\hspace{-0.2cm}/ \!-\! m^{0}) \psi
                       \!+\! \frac{G}{2} \sum_{a\!=\!0}^{8}
					                       \left[    \left( \bar{\psi} \lambda^{a}\psi \right)^{2}
											   \!+\! \left( \bar{\psi} i\gamma_{5} \lambda^{a}\psi \right)^{2}
										   \right] 
			           - \bar{\psi}M\theta(r\!-\!R)\psi,
\label{eq : NJL_bag}
\end{eqnarray}
where $\psi \!=\! (u, d, s)^{t}$ is the quark field and $m^{0} \!=\! \mbox{diag}(m_{u}^{0},m_{d}^{0},m_{s}^{0})$ current mass matrix.
The second term of Eq.~(\ref{eq : NJL_bag}) is the four-point quark interaction invariant under $U(N_{f})_{L} \times U(N_{f})_{R}$ symmetry, in which $\lambda^{a}$ ($a\!=\!0, \cdots, 8$) are the Gell-Mann matrices normalized by $\mbox{tr}\lambda^{a}\lambda^{b} \!=\! 2\delta^{ab}$.
In our formulation, we do not consider $U(1)_{A}$ breaking for simplicity.

The last term in Eq.~(\ref{eq : NJL_bag}) has been used in the MIT bag model to impose quark confinement \cite{Bogolioubov, Thomas, Hosaka}.
Assuming that the strangelet has a spherical shape, the step function $\theta(r\!-\!R)$ is introduced, where $R$ is the bag radius.
That term represents a quark mass term with $M$ for the exterior region and quarks are confined in the region $r\!<\!R$ by taking the limit $M\!\rightarrow\!\infty$. 
It is well known that the last term of Eq.~(\ref{eq : NJL_bag}) breaks chiral symmetry explicitly at the bag surface.
In order to recover chiral symmetry there, we need to introduce the chiral field (pion) which is coupled to the quarks at the bag boundary.
This leads to the condition of the chiral bag model, in which the pion cloud exists around the bag, and the vacuum structure is modified  in the bag due to the strong pion-quark coupling which is known as the chiral Casimir effect \cite{Hosaka_Toki}.
Although, this property associated with the pion field does not appear explicitly in our model lagrangian Eq.~(\ref{eq : NJL_bag}), the term of the NJL interaction in Eq.~(\ref{eq : NJL_bag}) is considered to be responsible for the pion induced property.

The parameters in Eq.~(\ref{eq : NJL_bag}), such as the coupling constant $G$, a three dimensional momentum cut-off $\Lambda$ (see Eq.~(\ref{eq : NJL_MIT_e}) below) and the current masses $m_{u}^{0}\!=\!m_{d}^{0}$ and $m_{s}^{0}$  are determined to reproduce the pion mass $m_{\pi}\!=\!0.139$ GeV, the pion decay constant $f_{\pi}\!=\!0.093$ GeV and the averaged mass of the nucleon and the delta $m_{N\!+\!\Delta}\!=\!1.134$ GeV.
 The mass is considered to be related with the dynamical quark mass in the vacuum by $m_{N\!+\!\Delta}\!\simeq\!3m_{u}^{\ast}$.
Then, we obtained the parameter set $G\Lambda^{2}\!=\!4.7$, $\Lambda\!=\!0.6$ GeV and $m_{u}^{0}\!=\!m_{d}^{0}\!=\!5.9\!\times\!10^{-3}$ GeV.
In this paper, we consider the current mass of the strange quark $m_{s}^{0}$ as a free parameter.
We show the results by setting $m_{s}^{0}\!=\!0.1$ GeV.
Other choices of $m_{s}^{0}$ do not affect our final conclusions qualitatively.

Now, let us investigate chiral symmetry breaking in a quark bag.
As usual, in the NJL interaction term in Eq.~(\ref{eq : NJL_bag}), we adopt the mean field approximation $(\bar{q}q)^{2} \!\rightarrow\! 2\bar{q}q\langle\bar{q}q\rangle \!-\! \langle \bar{q}q \rangle^{2}$, where $q \!=\!u, d$ and $s$, and solve the following gap equation
\begin{eqnarray}
 m_{q} \!=\! m_{q}^{0} \!-\! 2G \langle \bar{q}q  \rangle.
 \label{eq : meanfield}
\end{eqnarray}
Here we need to solve this equation for a quark bag. 
This requires a treatment of quark states in the quark bag with discretized energy levels, which is rather complicated.
In order to simplify the numerical calculations, we perform momentum integration with a density of states, which is approximately obtained by the multiple reflection expansion (MRE) \cite{MRE, Madsen}.
It is expressed by a smoothed function
\begin{eqnarray}
\rho_{MRE}(p,m,R) = 1+  \frac{6\pi^{2}}{pR} f_{S} (p/m) + \frac{12\pi^2}{(pR)^2} f_{C}(p/m) + \cdots,
\label{eq : MRE}
\end{eqnarray}
where $p$ is the momentum, $m$ the dynamical quark mass  and $R$ the radius of the quark bag.
The second and third terms in Eq.~(\ref{eq : MRE}) are the correction terms by the surface and curvature effects.
The functions $f_{S}$ and $f_{C}$ are given by  
\begin{eqnarray}
f_{S}(x)=-\frac{1}{8\pi} \left(1-\frac{2}{\pi} \arctan x \right),
\\ \nonumber
f_{C}(x)=\frac{1}{12\pi^2} \left[1-\frac{3x}{2}\left(\frac{\pi}{2}-\arctan x \right)\right].
\end{eqnarray}
In the limit $m\!\rightarrow\!0$, $f_{S}$ and $f_{C}$ become constants;
\begin{eqnarray}
\lim_{m\rightarrow0} f_{S}(p/m) \!=\! 0,
\\ \nonumber
\lim_{m\rightarrow0} f_{C}(p/m) \!=\! -\frac{1}{24\pi^{2}}.
\end{eqnarray}

By using the MRE method, the energy density $\epsilon$ in a strangelet with a radius $R$ is given as \cite{Kiriyama_Hosaka}, 
\begin{eqnarray}
\epsilon &=& \sum_{q=u,d,s}\left[ \frac{(m_{q}-m_{q}^{0})^{2}}{4G} - \nu\int^{\Lambda}_{p_{q}^{F}} \sqrt{p^{2}+m_{q}^{2}} \hspace{0.1cm}\rho_{MRE}(p,m_{q},R) \hspace{0.1cm} \frac{p^{2}dp}{2\pi^{2}} \right] - \epsilon_{0},
\label{eq : NJL_MIT_e}
\end{eqnarray}
where the integral is modified by the density of state Eq.~(\ref{eq : MRE}).
Note that the dynamical quark mass $m_{q}$ in Eq.~(\ref{eq : NJL_MIT_e}) is determined self-consistently by Eq.~(\ref{eq : meanfield}) as stated below.
In Eq.~(\ref{eq : NJL_MIT_e}), $\nu\!=\!N_{spin}\!\times\!N_{color}\!=\!6$ is the degrees of degeneracy of spin and color, and $\Lambda$ in the integral is a three dimensional momentum cut-off.
The value $p_{q}^{F}$ is the Fermi momentum which is determined by
\begin{eqnarray}
\nu \int_{0}^{p_{q}^{F}} \rho_{MRE}(p,m_{q},R) \hspace{0.1cm} \frac{p^{2}dp}{2\pi^{2}} = n_{q},
\label{eq : fermi_momentum}
\end{eqnarray}
for a given quark number density  $n_{q}$ for each flavor $q\!=\!u, d$ and $s$.
In Eq.~(\ref{eq : NJL_MIT_e}), the last term $\epsilon_{0}$ is the energy density in the chirally broken vacuum of infinite volume
\begin{eqnarray}
\epsilon_{0} = \sum_{q=u,d,s}\left[ \frac{(m_{q}^{\ast}-m_{q}^{0})^{2}}{4G} - \nu\int_{0}^{\Lambda} \sqrt{p^{2}+m_{q}^{\ast 2}} \hspace{0.1cm} \frac{p^{2}dp}{2\pi^{2}}\right],
\end{eqnarray}
where $m_{q}^{\ast}$ is the dynamical quark mass in the vacuum.
Note that the energy density Eq.~(\ref{eq : NJL_MIT_e}) is written as a sum of the kinetic energy of the valence quarks and the effective bag constant just as in the MIT bag model,
\begin{eqnarray}
\epsilon \!=\! \sum_{q=u,d,s} \nu\int_{0}^{p_{q}^{F}} \sqrt{p^{2}+m_{q}^{2}} \hspace{0.1cm} \rho_{MRE}(p,m_{q},R) \hspace{0.1cm} \frac{p^{2}dp}{2\pi^{2}} \!+\! B_{eff},
\label{eq : NJL_MIT_e2}
\end{eqnarray}
where the effective bag constant is defined by
\begin{eqnarray}
B_{eff} \!=\! \sum_{q=u,d,s} \left[ \frac{(m_{q}-m_{q}^{0})^{2}}{4G}-\nu\int_{0}^{\Lambda} \sqrt{p^{2}+m_{q}^{2}} \hspace{0.1cm} \rho_{MRE}(p,m_{q},R) \frac{p^{2}dp}{2\pi^{2}} \right]
- \epsilon_{0}.
\end{eqnarray}
Note that the effective bag constant $B_{eff}$ depends on the quark density, which is different from the constant $B$ in the MIT bag model.
Then, by taking  $\partial \epsilon/\partial m_{q}\!=\!0$, the gap equation Eq.~(\ref{eq : meanfield}) is written as;
\begin{eqnarray}
m_{q} \!=\! m_{q}^{0} + 2G\nu\frac{\partial}{\partial m_{q}} \int_{p_{q}^{F}}^{\Lambda} \sqrt{p^{2} \!+\! m_{q}^{2}} \hspace{0.1cm}\rho_{MRE}(p, m_{q}, R) \frac{p^{2}dp}{2\pi^{2}}.
\label{eq : SDeq}
\end{eqnarray}
We mention that that the dynamical quark mass depends not only on the Fermi momentum, but also on the quark bag radius $R$.
It is a characteristic feature of a finite size system.

Now, by the energy density Eq.~(\ref{eq : NJL_MIT_e2}), the total energy of the strangelet with a radius $R$ is given by
\begin{eqnarray}
E = \epsilon \hspace{0.1cm}V \!+\! E_{c} \!-\! \frac{\alpha}{R},
\label{eq : NJL_MIT_e_total}
\end{eqnarray}
where $V\!=\!(4\pi/3)R^{3}$ is the volume of the strangelet.
By assuming a uniform charge distribution in the strangelet, the Coulomb energy $E_{c}$ is given by
\begin{eqnarray}
E_{c} \simeq \frac{3}{5}\frac{e^{2}Q^{2}}{R},
\label{eq : Coulomb}
\end{eqnarray}
where the total electric charge is given by $Q\!=\!\frac{2}{3}N_{u} \!-\! \frac{1}{3}N_{d} \!-\! \frac{1}{3}N_{s}$ with $N_{q}$ being the number of each quark $q\!=\!u, d$ and $s$.
The last term of Eq.~(\ref{eq : NJL_MIT_e_total}) is a phenomenological zero point energy of the bag ($\alpha\!\simeq\!2.04$).

We obtain the energy of strangelet in the following way.
First, we give a baryon number $A$ and a strangeness fraction $r_{s}\!=\!N_{s}/(N_{u}\!+\!N_{d}\!+\!N_{s})$ with $N_{u}\!=\!N_{d}$.
Then, for several radii $R$, we solve the gap equation Eq.~(\ref{eq : SDeq}) and obtain the dynamical quark mass $m_{q}$ in the cavity, which is a function of the radius $R$.
Then we find the minimum of the energy Eq.~(\ref{eq : NJL_MIT_e_total}) with respect to the radius $R$.

\section{Numerical result}

\subsection{Chiral restoration in a cavity}

The confinement term in our model lagrangian Eq.~(\ref{eq : NJL_bag}) is responsible for the effects of the finite volume of the strangelet.
In this subsection, we investigate dynamical chiral symmetry breaking in the empty cavity without valence quarks.

\begin{figure}[tbp]
\begin{minipage}{8cm}
\vspace*{0.0cm}
\centering
\includegraphics[width=8cm]{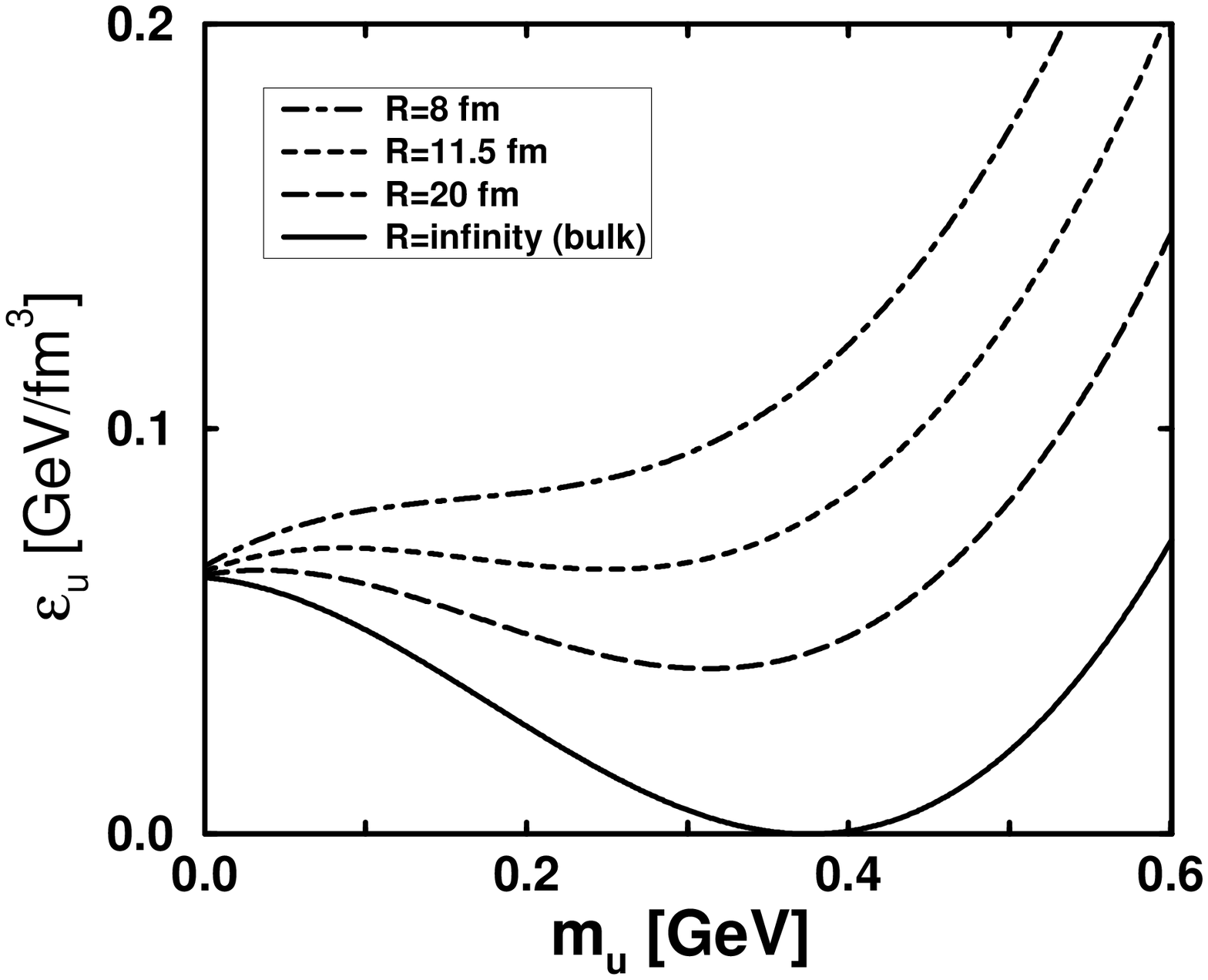}
\vspace{-0.0cm}
\end{minipage}
\begin{minipage}{8cm}
\centering
\includegraphics[width=8cm]{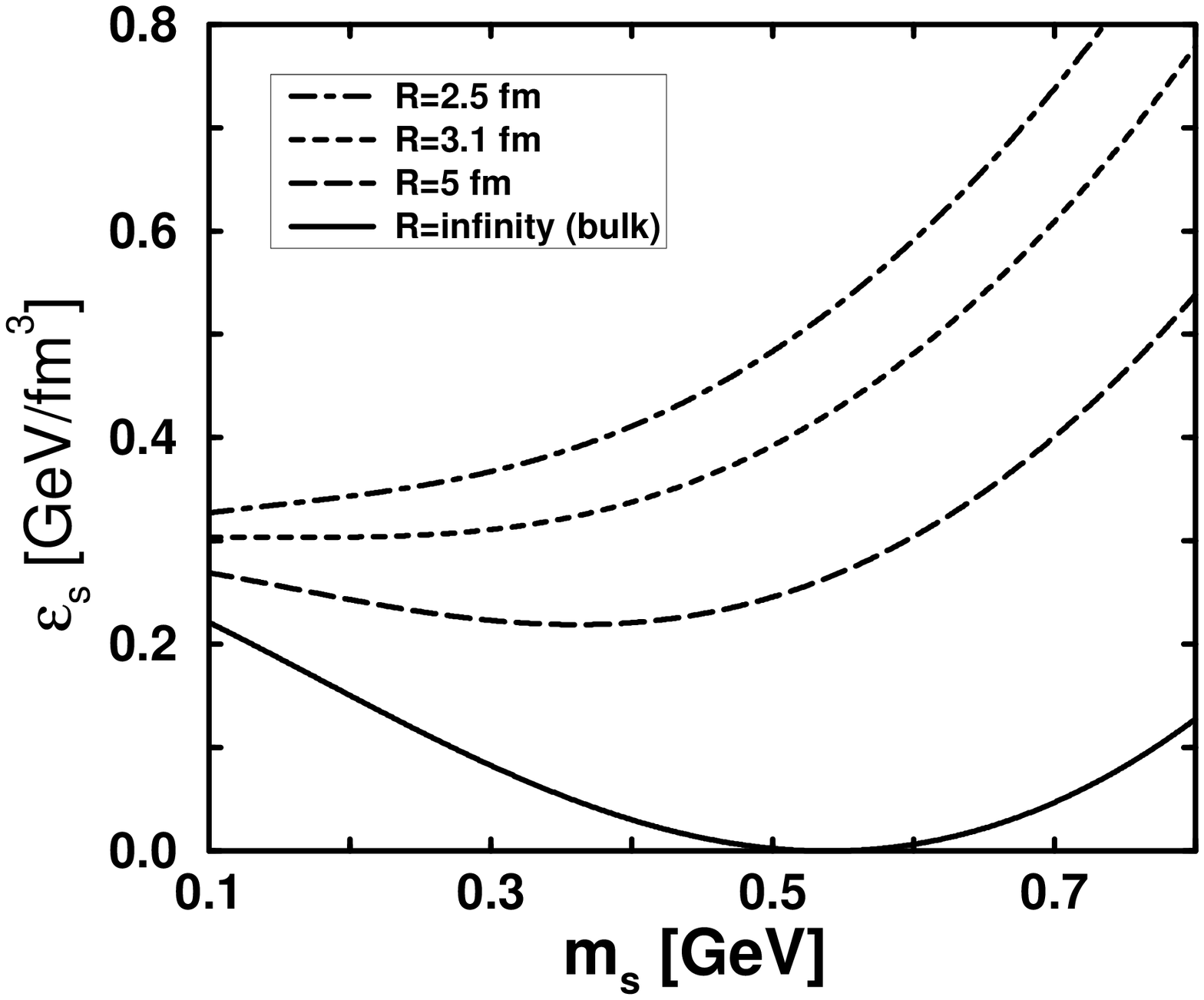}
\vspace{-0cm}
\end{minipage}
\caption{\small \baselineskip=0.5cm The energy density $\epsilon_{q}$ as a function of the dynamical quark mass $m_{q}$ for various radius $R$. Left: $u$ quark, right: $s$ quark.}
	\label{fig : e_m_m0mu0_cn_vacuum_R}
\end{figure}

In Fig.~\ref{fig : e_m_m0mu0_cn_vacuum_R}, we show the energy density  Eq.~(\ref{eq : NJL_MIT_e}) as a function of the dynamical quark mass in the empty cavity with several radii for $ud$ and $s$ quarks (dashed lines). 
For comparison,
the energy density of the bulk vacuum without the boundary condition is also plotted in  the same figure (solid lines).
Minimum points of the energy density provide the dynamical quark masses.
Concerning the $ud$ quark sector in the bulk vacuum, we obtain the dynamical quark mass $m_{u}\!=\!0.378$ GeV.
On the other hand, in the cavity, the dynamical quark masses are $m_{u}\!=\!0.322$, $0.258$ and $5.9\!\times\!10^{-3}(\!=\!m_{u}^{0})$ GeV for $R\!=\!20$, $11.5 \mbox{~and~} 8$ fm, respectively.
We see that the dynamical quark mass becomes smaller as the radius decreases.
This shows that chiral symmetry in the cavity tends to be restored.
It is also true for the $s$ quark sector.
In the bulk vacuum, we obtain the dynamical quark mass $m_{s}\!=\!0.539$ GeV, while we find $m_{s}\!=\!0.361$, $0.175$ and $0.1(\!=\!m_{s}^{0})$ GeV for the radius $R\!=\!5.0$, $3.1$ and $2.5$ fm, respectively.

\begin{figure}[tbp]
    \begin{center}
        \includegraphics[width=8cm, height=8cm, keepaspectratio, clip]{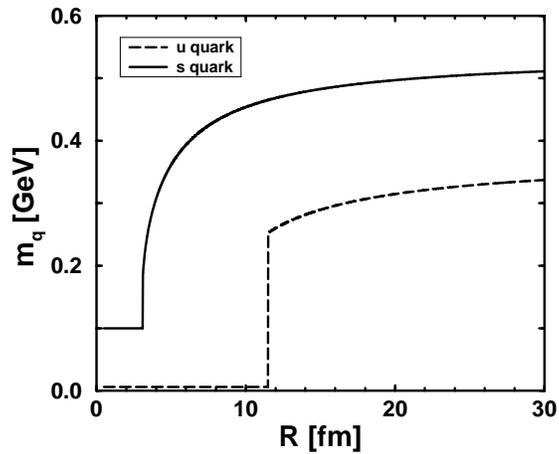}
    \end{center}
	\vspace*{-1.0cm}
	\caption{\small \baselineskip=0.5cm The dynamical quark mass $m_{q}$ of the $u$ and $s$ quarks as a function of the radius $R$.}
	\label{fig : m_R_m0_cn_vacuum}
\end{figure}

In Fig.~\ref{fig : m_R_m0_cn_vacuum}, the dynamical quark mass is shown as a function of the radius of the cavity.
The chiral symmetry in the $ud$ quark sector is restored at $R\!=\!11.5$ fm, while that in the $s$ quark sector is restored at $R\!=\!3.1$ fm.
We see that the chiral restoration in the $s$ quark is suppressed as compared with the $ud$ quark.
This is considered to be due to the large current mass of the $s$ quark as compared with the $ud$ quark.
The difference in the tendency of the chiral symmetry restoration in the $ud$ quark sector and the $s$ quark sector causes the difference in the stability of the $ud$ quark droplets and the strangelets.

\subsection{Stability of strangelet}

Now we discuss the stability of the strangelet.
For this purpose, we add $3A$ valence quarks in the cavity, and calculate the energy per baryon number $E/A$ for several baryon numbers $A$ and the strangeness fractions $r_{s}$.
In order to simplify the discussions, first, we fix the strangeness fraction $r_{s}\!=\!0$ for the $ud$ quark droplets and $r_{s}\!=\!1/3$ for the strangelets, respectively.
In the case of $r_{s}\!=\!1/3$, the Coulomb energy of Eq.~(\ref{eq : Coulomb}) vanishes.
In this subsection, we turn off the Coulomb term.

\begin{figure}[tbp]
\begin{minipage}{8cm}
\vspace*{0.0cm}
\centering
\includegraphics[width=8cm]{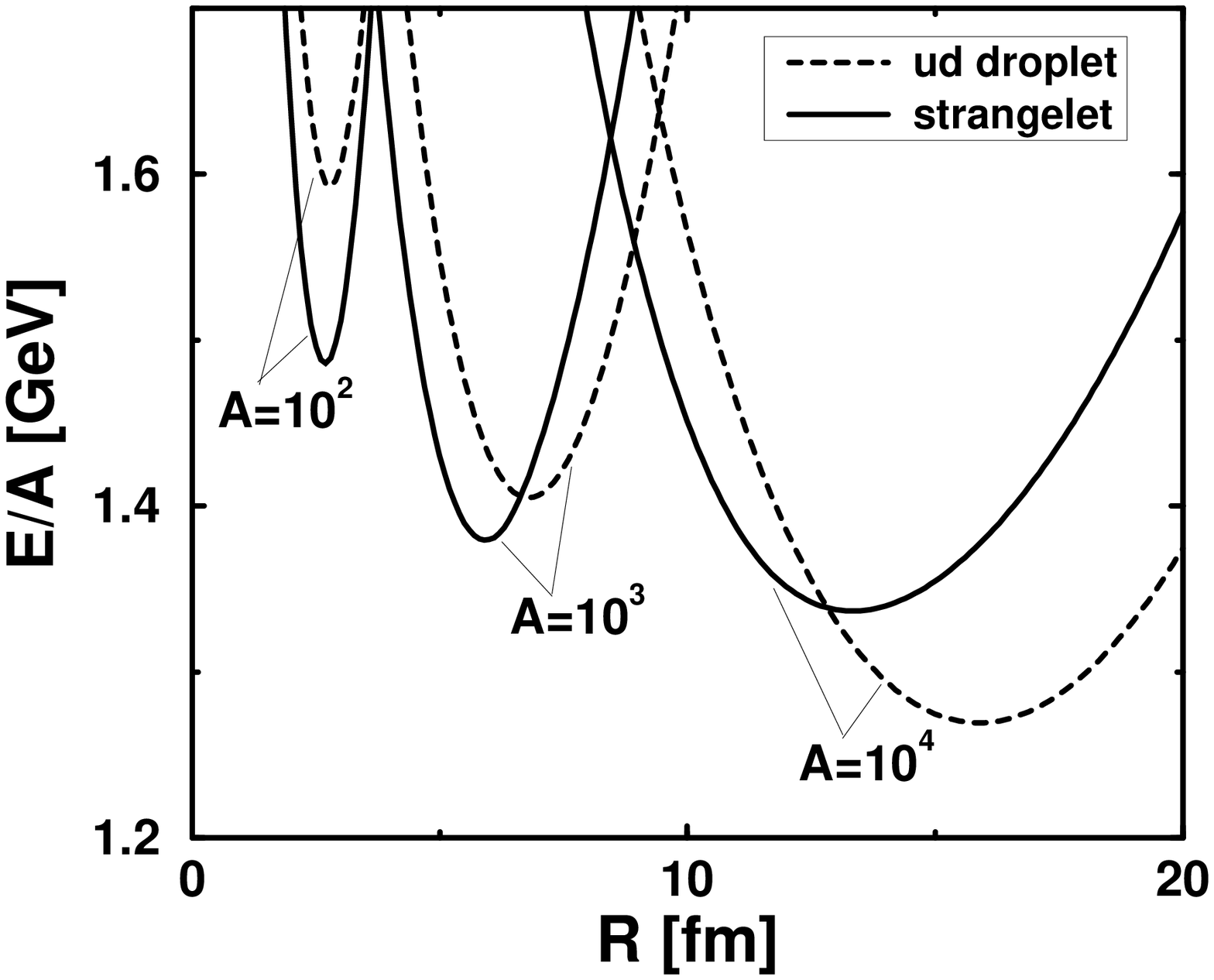}
\vspace{-0.0cm}
\end{minipage}
\begin{minipage}{8cm}
\centering
\includegraphics[width=8cm]{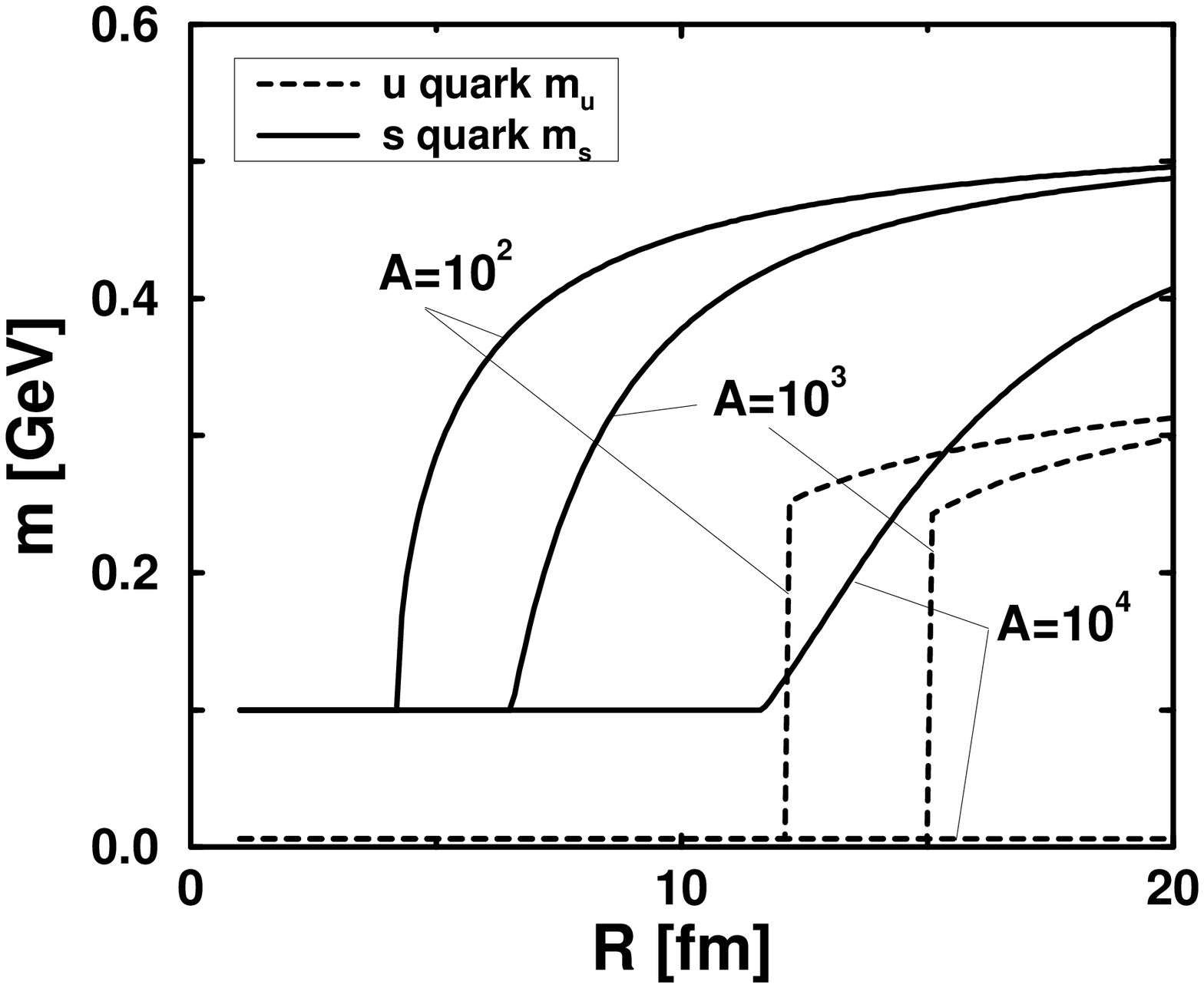}
\vspace{-0cm}
\end{minipage}
\caption{\small \baselineskip=0.5cm Left: (a) The energy per baryon number $E/A$ as a function of the radius $R$ for the $ud$ quark droplet ($r_{s}\!=\!0$) and the strangelet ($r_{s}\!=\!1/3$). Right: (b) The dynamical quark mass $m_{q}$ of the $ud$ and $s$ quarks in the strangelets with $r_{s}\!=\!1/3$.}
	\label{fig : E_R_rs_A}
\end{figure}

The cavity radius $R$ of a strangelet is determined by the variation of $E/A$.
In Fig. \ref{fig : E_R_rs_A}(a), we show the energy per baryon number $E/A$ as a function of the cavity radius $R$
 for the  baryon numbers $A\!=\!10^{2}, 10^{3}$ and $10^{4}$.
The minimum of $E/A$ gives the energy and the radius of the strangelet.
The existence of the minimum is understood in the following way.
In the total energy Eq.~(\ref{eq : NJL_MIT_e_total}) and (\ref{eq : NJL_MIT_e2}), there are two terms: the kinetic energy from the valence quarks and the volume energy $B_{eff}V$ from the effective bag constant.
As the radius becomes large, the kinetic energy decreases, while the volume energy increases.
On the other hand, as the radius becomes small, the kinetic energy increases, while the volume energy decreases.
Then, we find the equilibrium radius balanced by the kinetic energy and the volume energy.

For example, for the baryon number $A\!=\!10^{4}$, the energy per baryon number in the $ud$ quark droplet and the strangelet are $(E/A)_{ud}\!=\!1.27$ GeV and $(E/A)_{uds}\!=\!1.33$ GeV, respectively.
For the baryon number $A\!=\!10^{2}$, the energy per baryon number in the $ud$ quark droplet and the strangelet are $(E/A)_{ud}\!=\!1.60$ GeV and $(E/A)_{uds}\!=\!1.48$ GeV, respectively.
In our results, the strangelets are more stable than the $ud$ quark droplets for smaller baryon numbers $A\!<\!2\!\times\!10^{3}$.
The stability of the strangelets with small baryon numbers is very much different from the result for the bulk quark matter, where the strange matter with infinite volume is not absolutely stable \cite{Buballa96, Buballa98, Buballa99}.
This is because of the effect of the confinement leading to the restoration of chiral symmetry in the cavity. 
In order to show the restoration of chiral symmetry in the strangelets
 in Fig. \ref{fig : E_R_rs_A}(b), the dynamical quark masses $m_{u}$ and $m_{s}$ of the $ud$ and $s$ quarks are shown as functions of the cavity radius $R$.
We see that chiral symmetry of the $ud$ quark and $s$ quark in the strangelet has a tendency to be restored for small radii as seen in the empty cavity.

\begin{figure}[tbp]
\begin{minipage}{8cm}
\vspace*{0.0cm}
\centering
\includegraphics[width=8cm]{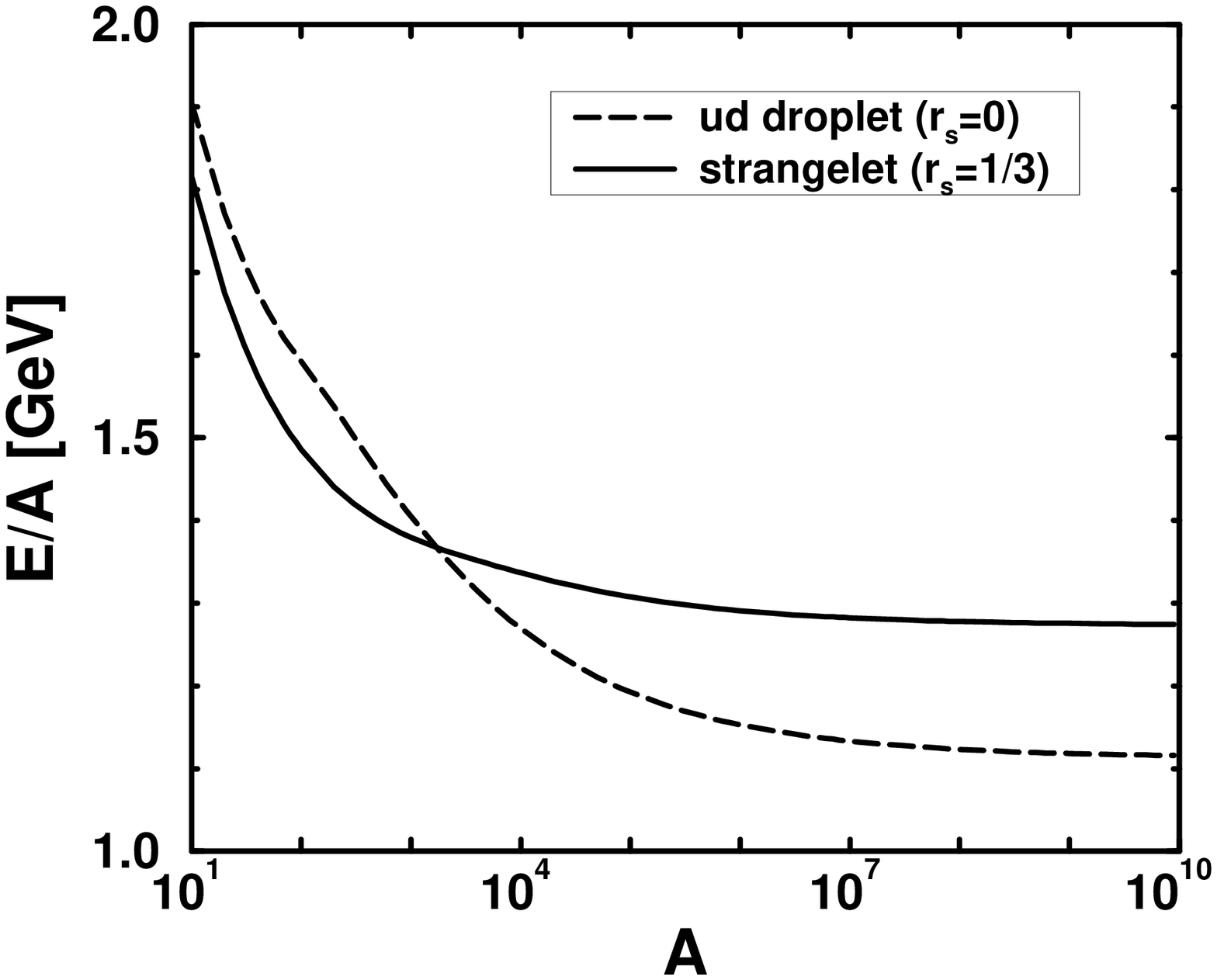}
\vspace{-0.0cm}
\end{minipage}
\begin{minipage}{8cm}
\centering
\includegraphics[width=8cm]{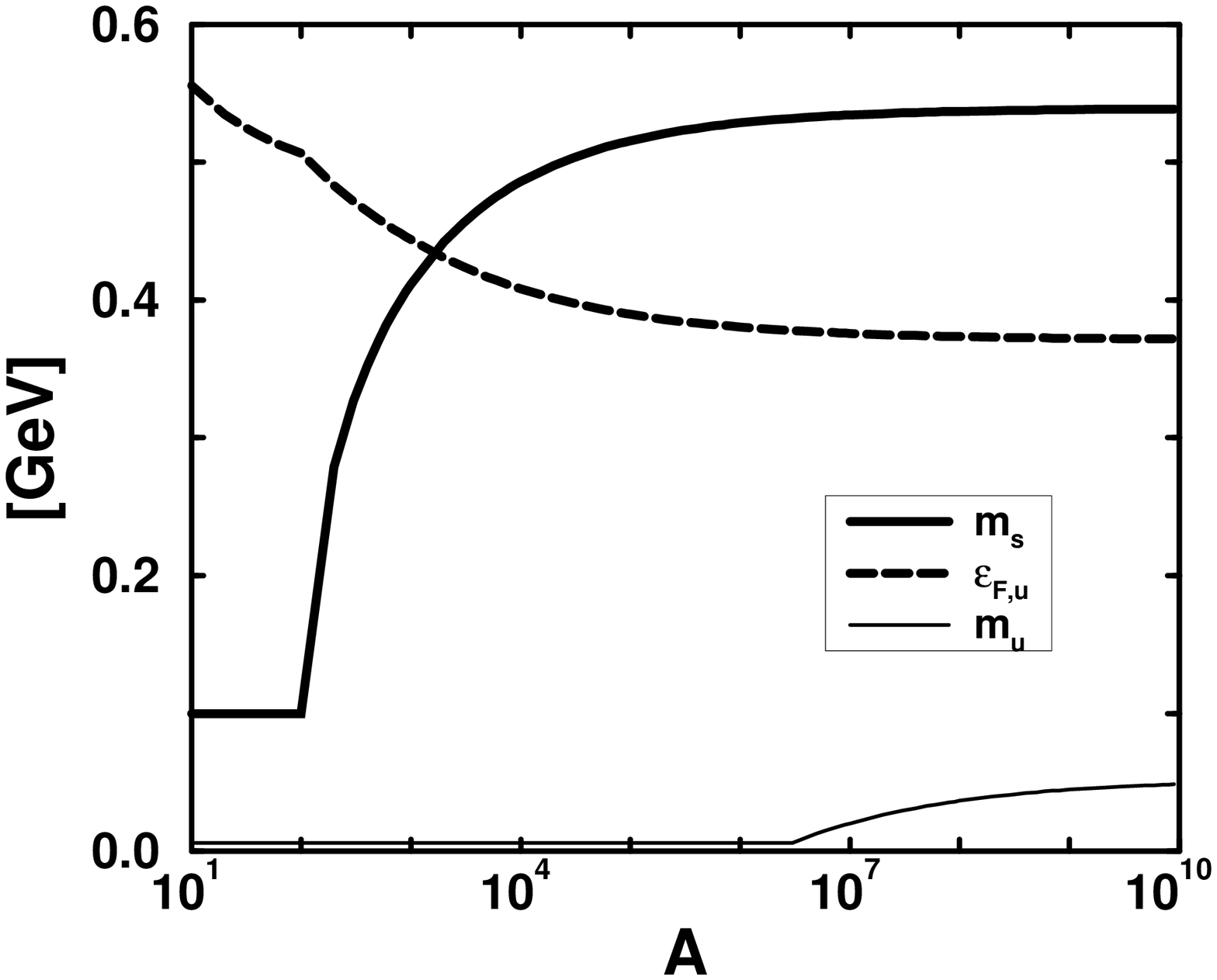}
\vspace{-0cm}
\end{minipage}
\caption{\small \baselineskip=0.5cm Left:  (a) The energy per baryon number $E/A$ as a function of the baryon number $A$. The solid line for the strangelets ($r_{s}\!=\!1/3$) and the dashed line for the $ud$ quark droplets ($r_{s}\!=\!0$). Right: (b) The dynamical quark mass $m_{u}$ and $m_{s}$ of the $u$ and $s$ quarks, and the Fermi energy $\epsilon_{F,u}$ of the $u$ quark in the $ud$ quark droplets with $r_{s}\!=\!0$.}
	\label{fig : E_A_ms0100_cn_rs}
\end{figure}

In Fig.~\ref{fig : E_A_ms0100_cn_rs}(a), we show explicitly the energy per baryon number $E/A$ as a function of the baryon number $A$ for $ud$ quark droplets and strangelets.
It is shown that the strangelets are more stable than the $ud$ quark droplets for the baryon number $A\!<\!2\!\times\!10^{3}$.
It is generally expected that the strange matter can be more stable than the $ud$ quark matter, when the dynamical quark mass of $s$ quark $m_{s}$ is smaller than the Fermi energy  $\epsilon_{F,u}$ of the $ud$ quark.
When this relation is satisfied, the weak transition from $ud$ quarks to $s$ quarks can occur by the weak processes $u \!\rightarrow\! d \!+\! e^{+} \!+\! \nu_{e}$ and $u \!+\! d \!\rightarrow\! u \!+\! s$.
In Fig.~\ref{fig : E_A_ms0100_cn_rs}(b), we compare $m_{s}$ and $\epsilon_{F,u}$ in the $ud$ quark droplets.
It is shown that $m_{s} \!<\! \epsilon_{F,u}$ is satisfied in the quark droplets with $A\!<\!2\!\times\!10^{3}$.

So far, we have discussed the stability of the strangelets as compared with the $ud$ quark droplets.
We have found that the absolute values of the energy of the strangelet for $A\!<\!2\!\times\!10^{3}$ is larger than the masses of normal nuclei.
Consequently, the strangelets may not be absolutely stable as the QCD matter.
However, if there are many $s$ quarks in the strangelets, the decay of the strangelets to the normal nuclei would take a long time  due to the many steps in the  weak processes.
Therefore the strangelets could survive as stable particles, once they are formed.

\subsection{Observables}

When we consider that the strangelets are formed in the QCD phase transition in the early universe and/or in the explosions of the strange stars, the remaining strangelets could be observed in the cosmic rays.
They would be observed as exotic particles with a large mass and a small electric charge.
In order to identify such heavy particles, the charge-to-baryon number ratio is an important quantity.
For this purpose, let us concentrate on the stable strangelets of $A\!<\!2\!\times\!10^{3}$.

\begin{figure}[tbp]
\begin{minipage}{8cm}
\vspace*{0.5cm}
\includegraphics[width=8cm]{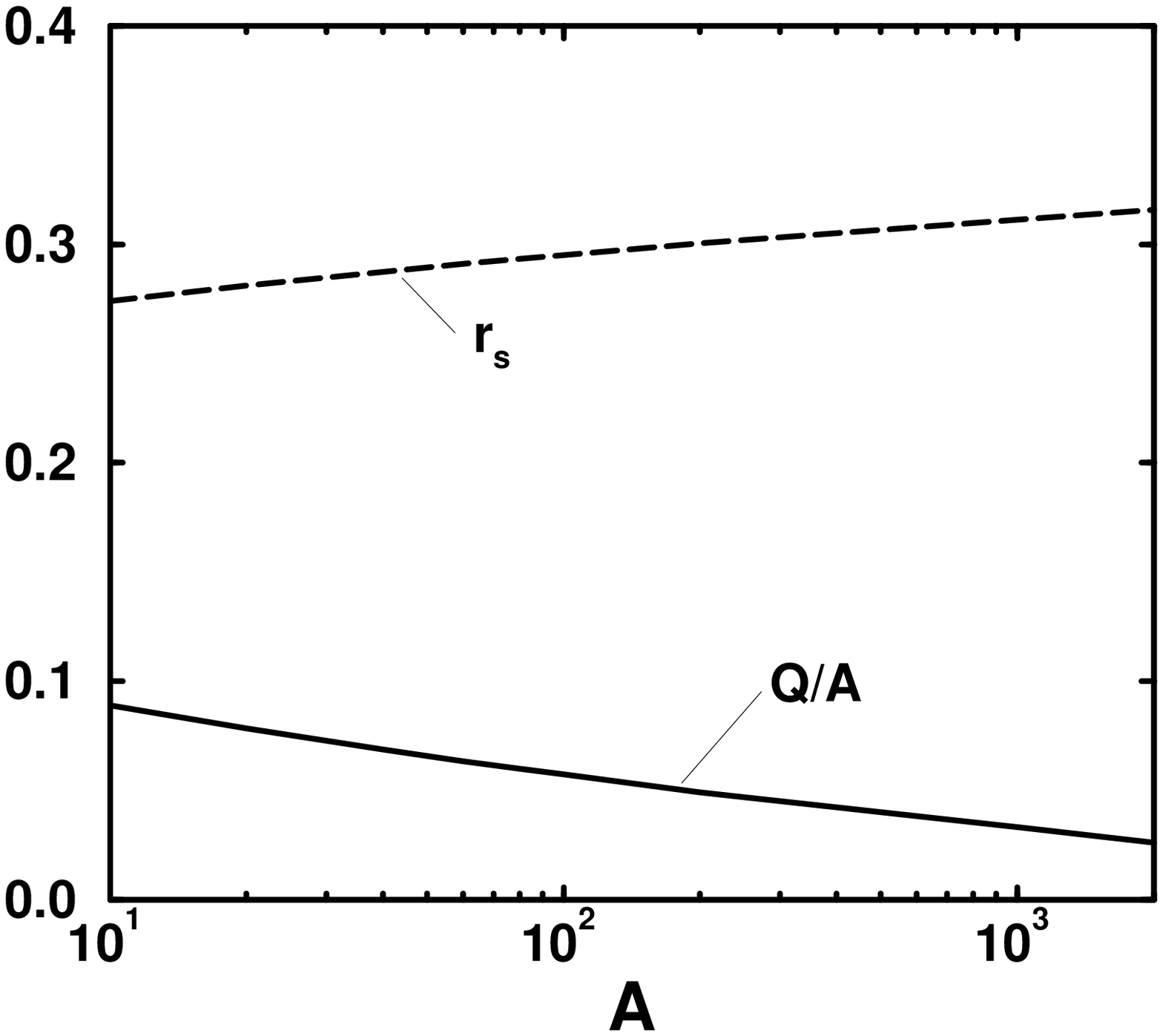}
\caption{\small \baselineskip=0.5cm The strangeness fraction $r_{s}$ and the electric charge $Q/A$ of the strangelets as a function of the baryon number $A$.}
	\label{fig : QperA_A_ms0100_cy}
\end{minipage}
\begin{minipage}{8cm}
\includegraphics[width=8cm]{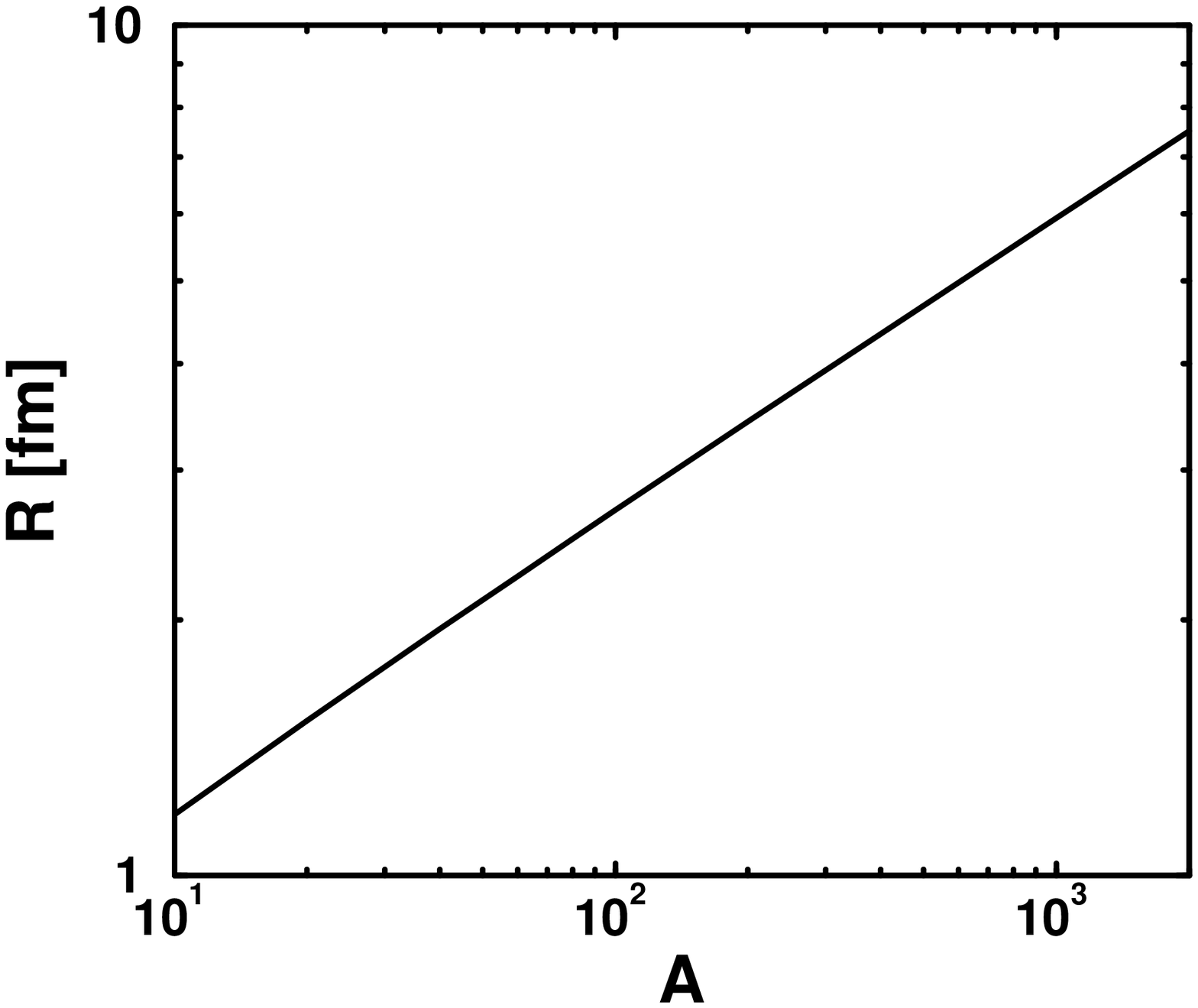}
\caption{\small \baselineskip=0.5cm The radius $R$ of the strangelet as a function of the baryon number $A$.}
	\label{fig : Rf_A_ms0100_cy}
\end{minipage}
\hspace{1cm}
\end{figure}

In the previous subsection,
 we have fixed the ratio of $s$ quarks to be $r_{s}\!=\!0$ and $1/3$.
In the following, we consider a variation of the energy Eq.~(\ref{eq : NJL_MIT_e_total}) with respect to $r_{s}$.
The resulting strangeness fraction $r_{s}$ is plotted  as a function of the baryon number $A$ in Fig.~\ref{fig : QperA_A_ms0100_cy}.
We see that the strangeness fraction of the strangelets with $A\!<\!2\!\times\!10^{3}$ is close to $r_{s}\!=\!1/3$.

Once the strangeness fraction $r_{s}$ is obtained, the number of quarks $N_{q}$ for $q\!=\!u, d$ and $s$
 are also obtained.
Then, the electric charge $Q$ of the strangelets are calculated from $Q\!=\!\frac{2}{3}N_{u}\!-\!\frac{1}{3}N_{d}\!-\!\frac{1}{3}N_{s}$. 
In Fig.~\ref{fig : QperA_A_ms0100_cy}, we show also the charge-to-baryon number ratio $Q/A$ as a function of the baryon number $A$.
The electric charge of the strangelets are of order a few percents of the normal nuclei.
Such strangelets can have large baryon number, since the Coulomb instability can be negligible. 
Thus, our results show that the strangelets would be the exotic particles with small charge-to-baryon number ratio as compared with the normal nuclei.

Let us compare our theoretical results and the existing data which was reported in the observation of exotic particles in the cosmic rays.
We show the baryon number and the charge-to-baryon number ratio of the observed particles in Table \ref{tbl : Table1}.
The baryon numbers $A$ of these exotic particles are from the order of hundreds to thousands.
The charge-to-baryon number $Q/A$ is around 0.04.
It is interesting that these observed values are very close to our  theoretical results.
These exotic particles could be candidate of strangelets.

In Fig.~\ref{fig : Rf_A_ms0100_cy}, we show the radius $R$ of the strangelets as a function of the baryon number $A$.
The relation between $R$ and $A$ is expressed approximately by
\begin{eqnarray}
R =  r_{0}A^{1/3},
\end{eqnarray}
with $r_{0}\!\simeq\! 0.57$ fm.
Then, the baryon number density in the strangelet is estimated as $n_{B}\!=\!A/(4\pi R^{3}/3)\!=\!7.6\;n_{B}^{0}$ with the normal nuclear matter density $n_{B}^{0}\!=\!0.17$ $\mbox{fm}^{-3}$.
Recently, the $\bar{K}$-nuclei including the strangeness is discussed to have such a high density \cite{Akaishi, AkaishiDote}.
Though the relation between the strangelets and the $\bar{K}$-nuclei is not yet clear, it is interesting that such high baryon number density in the strangelets is comparable with the result in the $\bar{K}$-nuclei.

So far, we did not include electrons in our discussions.
We show that the electrons play only minor role in the strangelets.
From Fig.~\ref{fig : Rf_A_ms0100_cy}, the electric charge of the strangelets are at most $Q\!\simeq\!55$ for $A\!=\!2\!\times\!10^{3}$, due to small $Q/A\!\simeq\!0.026$.
Assuming an electron around the strangelets with such electric charge, the de Broglie wave length of the electron is estimated by the energy of the electron 
$
E_{e} = \sqrt{p^{2}+m_{e}^{2}} - e^{2}Q/r,
$
where $p$ is the momentum of the electron, $m_{e}\!=\!0.51$ MeV the mass of the electron and $r$ the distance from the strangelet.
Then, we find that the de Broglie wave length of the electron is $\lambda\!\simeq\!960$ fm for $A\!=\!2\!\times\!10^{3}$.
That is much longer than the radius of the strangelet, which is shown in Fig.~\ref{fig : Rf_A_ms0100_cy}.
Therefore, the electrons exist almost outside of the strangelet.
If the size of the strangelets could be larger than the electron de Broglie wave length, we need to consider the electrons in the strangelets.

\begin{table}[htdp]
\begin{center}
\begin{tabular}{|c|c|c|}\hline
baryon number $A$ & charge-to-baryon number $Q/A$ &  \\
\hline
$A\!\sim\!350\!-\!450$ & $0.03\!-\!0.04$  &  \cite{Kasuya} \\
$A\!\sim\!460$ & $0.043$  &  \cite{Ichimura} \\
$A\!>\! 1000$ & $0.046$  & \cite{Price} \\
$A\!\sim\!370$ & $0.038$  & \cite{Miyamura, Capdevielle} \\
\hline
\end{tabular}
\end{center}
\caption{\small \baselineskip=0.5cm The baryon number $A$ and the charge-to-baryon number $Q/A$ from cosmic ray experiments. See also \cite{Banerjee}.}
\label{tbl : Table1}
\end{table}

\section{Conclusion}
We discussed the stability of the strangelet by considering dynamical chiral symmetry breaking.
We used the NJL interaction for chiral symmetry breaking.
In addition, we incorporated the spherical cavity for the confinement of quarks by the boundary condition approximately.
In the mean field approximation in the finite volume system, we obtained the gap equation for the dynamical generation of the quark mass.
Then, we obtained the energy of the strangelets.

As a result, it was shown that chiral symmetry tended to be restored in the cavity at small radii.
Then, the dynamical quark mass became small as compared with that in the vacuum of infinite volume,  and the strange quark mass can be smaller than the Fermi energy of the $ud$ quarks in the droplets. 
We investigated the stability of the strangelet for several baryon number $A$ and the strangeness fraction $r_{s}\!=\!0$ and $1/3$.
Then, it was shown that the strangelets are more stable than the $ud$ quark droplets for the baryon number $A\!<\!2\!\times\!10^{3}$ for the $s$ quark current mass $m_{s}^{0}\!=\!0.1$ GeV.
Our result did not change qualitatively for the case of $m_{s}^{0}\!=\!0.18$ GeV, in which we obtained the stable strangelets for $A\!<\!0.5\!\times\!10^{3}$ .
We obtained the charge-to-baryon number of the strangelets, which is consistent with the data which was reported to be observed in the heavy particles in the cosmic rays.

Finally, we comment on the further developments in the study of the strangelets.
The energy per baryon number $E/A$ in the stable strangelets with $A\!<\!2\!\times\!10^{3}$ seems to have  larger energy than ordinary nuclei.
For the more realistic discussions, the color exchange interaction between dynamical quarks should be introduced.
It is known in the quark model that the color exchange interaction splits the mass of the nucleon $N$ and the delta $\Delta$, while, in our calculation, we considered that the masses of the $N$ and $\Delta$ are degenerate.
Furthermore, the quark-quark pairing, namely the color superconductivity, in the quark droplets may affect the energy of the strangelet \cite{Madsen01, Amore, Kiriyama}.
Especially, it is discussed that the strange matter can form the color-flavor locked phase.
Such a condensed state could play an important role in the strangelets.

\vspace{0.5cm}
We would like to express our gratitude to  Prof. S.~Raha and  Prof. N.~Sandulescu for interesting suggestions to start our study of  strangelets.
We would express thanks to Prof. Itahashi for fruitful suggestions from the view of experiments. 
We wish  to thank Dr. O.~Kiriyama for many discussions from an early stage of our study.


\end{document}